# Introducing an ensemble method for the early detection of Alzheimer's disease through the analysis of PET scan images

Arezoo Borji[1], Taha-Hossein Hejazi [1*], and Abbas Seifi[2]

[1]Department of Industrial Engineering, College of Garmsar, Amirkabir University of Technology (Tehran Polytechnic), Tehran, Iran
[2]Department of Industrial Engineering and Management Systems, Amirkabir University of Technology (Tehran Polytechnic), Tehran, Iran

**Abstract**

Alzheimer's disease is a progressive neurodegenerative disorder that primarily affects cognitive functions such as memory, thinking, and behavior. In this disease, there is a critical phase, mild cognitive impairment, that is really important to be diagnosed early since some patients with progressive MCI will develop the disease. This study delves into the challenging task of classifying Alzheimer's patients into four distinct groups: control normal (CN), progressive mild cognitive impairment (pMCI), stable mild cognitive impairment (sMCI), and Alzheimer's disease (AD). This classification is based on a thorough examination of PET scan images obtained from the ADNI dataset, which provides a thorough understanding of the disease's progression. Several deep-learning and traditional machine-learning models have been used to detect Alzheimer's disease. In this paper, three deep-learning models, namely VGG16 and AlexNet, and a custom Convolutional neural network (CNN) with 8-fold cross-validation have been used for classification. Finally, an ensemble technique is used to improve the overall result of these models. The classsofication results show that using deep-learning models to tell the difference between MCI patients gives an overall average accuracy of 93.13% and an AUC of 94.4%.

## 1. Introduction

Alzheimer's disease represents a significant global challenge in the healthcare sector, [1] placing it among the top concerns worldwide [2]. A critical stage in this disease is mild cognitive impairment (MCI), where early identification is crucial for those affected [3]. This stage is categorized into two types: stable MCI and progressive MCI. Identifying stable MCI allows for treatment, and early detection of progressive MCI can slow its advancement. Currently, various neuroimaging methods have been developed to aid in the challenging prediction of transitions between stable and progressive MCI [4]. In this study, we introduce a promising method based on voxel-based morphometry (VBM) analysis on PET-Scan image data for early prediction of Alzheimer. We survey how to extract features from sMCI and pMCI subjects based on patterns of gray matter. We use VGG16, AlexNet, and a custom convolutional neural network (CNN) to classify AD vs. CN and pMCI vs. sMCI. Finally, a method called "ensemble method utilizing majority voting " is used to combine the outputs of several 3D dense networks. The remainder of this paper is organized as follows; first, section 2 addresses related work in the scope of classification of Alzheimer's disease. In section 3, we present our methodology for the classification and image processing steps used for the dataset. Our computational results and their comparison to the related works are presented in section 4. Finally, the conclusion and future research are presented in section 5.

## 2. Literature review

In this part, we survey numerous research studies that have addressed the challenges of the classification of Alzheimer's disease. Naami et al. [5] confronted the high-dimensional problem of brain images by utilizing an artificial neural network (ANN) for classification. This approach adeptly navigated the intricate nature of the data, effectively distinguishing between cognitively normal (CN) and AD individuals. Mahmood et al. [6] proposed a hybrid methodology incorporating principal component analysis (PCA) for dimension reduction and ANN for classification. Accordingly, their method could discriminate between CN and AD individuals with good accuracy. Turning attention to diffusion tensor imaging (DTI) images, Kar et al. [7] utilized a fuzzy technique to discern between CN and AD subjects. Their approach was successful in handling the complexities of DTI data and achieving precise classification of samples. Veen et al. [8] looked into how well generalized matrix learning vector quantization (GMLVQ) could classify people with AD. This study shows how important it is to look at both global and local correlations for correct disease classification. Biomarkers that are not invasive, like conventional methods in medical image processing, typically use prior knowledge to segment brain images into Regions of Interest (ROI) [9] or Voxel-Based Morphometry (VBM) [10]. Deep-learning algorithms can discover hidden representations among multiple regions of neuroimages, these algorithms outperform conventional methods in identifying AD patterns. Addressing the challenge of selecting significant ROIs in whole-brain analysis, Ortiz et al. [11] introduced a novel strategy using self-organizing maps (SOM). However, this method necessitated prior knowledge for extracting features from specific brain ROIs. Gorji et al. [12] focused on feature extraction from MR images using image moments, while Jha et al. [13] proposed the Dual-Tree Complex Wavelet Transform (DTCWT) as an alternative and said it was better at choosing directions than the Discrete Wavelet Transform (DWT). Principal Component Analysis (PCA) emerged as a widely adopted technique for significantly reducing the dimensionality of the feature space. Lopez et al. [14] employed Bayesian classifiers for AD classification, while Horn et al. [15] utilized Partial Least Squares (PLS) regression to minimize ROI-based features. K-Nearest Neighbors (K-NN) emerged as the most accurate method for CN versus AD classification, indicating its potential for handling AD detection tasks. In the context of survival analysis on patients with head and neck squamous cell carcinoma, Leger et al. [16] explored various machine-learning techniques for feature selection. Hu et al. [17] employed an image dataset to construct a correlation matrix, classifying it using a targeted autoencoder network for AD detection, showcasing the potential of deep learning in neuroimaging analysis. She et al. introduced a multimodal stacked deep probabilistic network (MM-SDPN) for Alzheimer's diagnosis, demonstrating the efficacy of combining multiple data modalities to enhance AD classification accuracy. Suk et al. [18] solved the problem of getting relevant features out of large amounts of data and made AD detection better. In addition, it has been demonstrated that sparse regression models are helpful in managing high-dimensional data with a limited number of training samples [19], holding promise in addressing the challenges posed by limited data availability in AD detection and classification. In summary, the literature review underscores the diversity of methodologies and techniques employed in AD detection, encompassing a range of machine-learning algorithms and deep-learning architectures.

## 3. Methodology

In this paper, we utilize the original PET images that comprise 363 samples: 96 cognitively normal (CN) subjects and 176 individuals with mild cognitive impairment (MCI), among whom 89 patients have progressive MCI that is expected to progress to Alzheimer's disease and 87 patients have stable MCI that is expected to remain in this phase. Additionally, there are 91 records of Alzheimer's disease (AD) patients. For performing our first stage method, which reduces the number of features, PET scan images that have been processed using co-registration, normalization, and VBM serve as input for the custom CNN, VGG16, and AlexNet models that have been applied to processed images. The comprehensive methodology process is shown in Figure 2. This schematic outlines the sequential steps involved in the proposed methodology, starting from image preprocessing (including co-registration and normalization), through feature extraction and augmentation, to the final classification via the deep learning models (VGG16, AlexNet, and the custom CNN), culminating in the ensemble method with majority voting for diagnosis.

### 3.1. Image processing

In this research paper, we employ original PET scan images sourced from the ADNI dataset. These images are three-dimensional and measure 160 × 160 × 160 mm in size. Furthermore, the voxel sizes, which indicate the image's resolution, are configured at 1.5 × 1.5 × 1.5 mm. To provide a visual representation of the images contained within the dataset, figure 1 showcases a selection of images from each category. This figure likely presents a visual comparison of PET scan images from individuals across the four categories of interest: cognitively normal (CN), stable mild cognitive impairment (sMCI), progressive mild cognitive impairment (pMCI), and Alzheimer's disease (AD). The progression from CN to AD is visualized, highlighting the morphological changes in the brain that correspond to the advancement of the disease.

This dataset serves as an invaluable resource for conducting research on Alzheimer's disease. Its detailed three-dimensional imaging and its capacity to categorize data into distinct groups establish a robust foundation for studies aimed at investigating, diagnosing, or gaining a deeper understanding of the progression of neurodegenerative diseases over time.

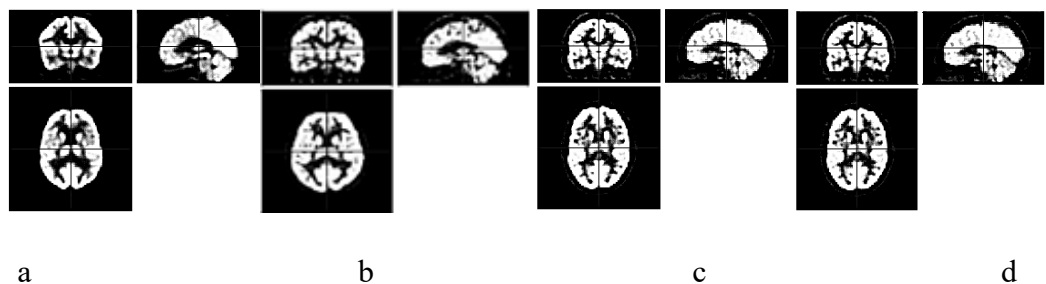

a  b  c  d

Fig. 1: Four records from each group of the dataset based on their severity [20].

Alzheimer's disease causes damage to brain cells in the gray matter primarily due to an accumulation of amyloid, resulting in the gradual loss of these cells [21]. Scientists employ voxel analysis, a comprehensive examination of brain regions, to gain insights into this process. Identifying the specific brain regions most impacted is essential for advancing our understanding of the disease. It's important to emphasize that for applying deep learning models, a well-balanced dataset is crucial to achieving accurate classification results [22]. Data augmentation methods, such as padding, rotation, and mirroring, are employed to generate multiple images from a single source, and equal numbers of samples are selected from every class within the dataset. Table 1 shows the information about the samples obtained for each class. This table likely provides quantitative details on the dataset, listing the number of subjects in each category (CN, pMCI, sMCI, AD) before and after data augmentation, along with other relevant information such as average age. This highlights the dataset's composition and the extent of augmentation applied to ensure a balanced and comprehensive training set for the models.

Table1. Information of subjects in each group before and after augmentation.

| Disease state | Number of subjects before augmentation | Number of subjects after augmentation | Age |
|---|---|---|---|
| CN | 96 | 8352 | 76.01±4.81 |
| pMCI | 89 | 8352 | 75.11±6.87 |
| sMCI | 87 | 8352 | 76.41±7.14 |
| AD | 91 | 8352 | 75.72±7.36 |

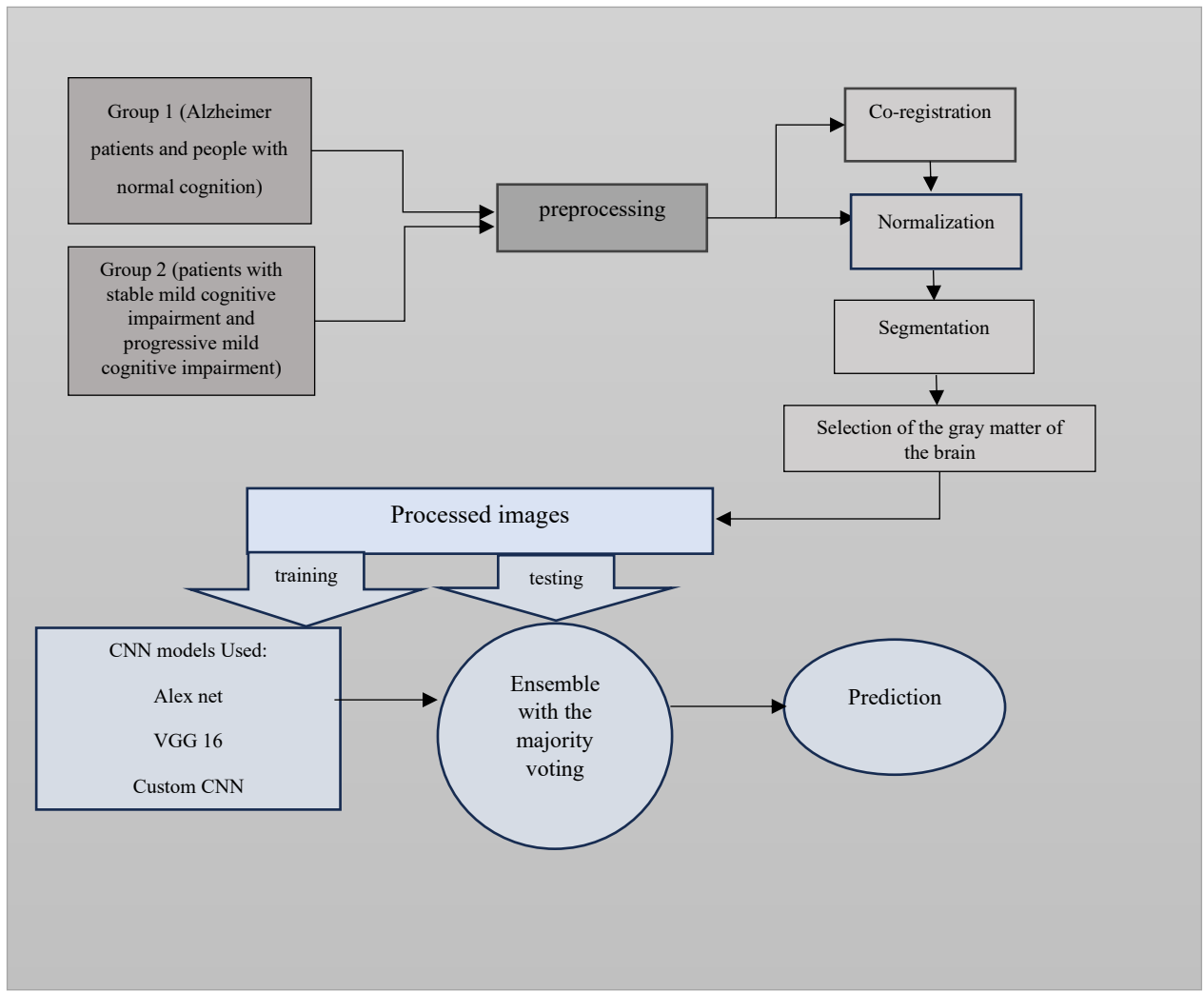

Fig. 2. The overall procedure of the proposed method.

### 3.1.1. Co-registration

Co-registration function is pivotal in ascertaining the compatibility of two or more scans [32]. It is essential to perform a thorough verification of co-registration before moving forward.

### 3.1.2. Normalization function

This function serves the crucial purpose of transferring scans into the standardized MNI ( Montreal Neurological Institute) space, as utilized in the SPM software. The standard coordinate system in the field of neuroimaging is the MNI space, which the Montreal Neurological Institute created from 152 scans. Given that brain images of the same individual, acquired during different sessions, may exhibit variations in the positioning of the head and brain, a common spatial reference is imperative for effectively comparing datasets from diverse individuals [23].

The process of transferring scans to the MNI space involves two essential steps: Bias correction: This initial step addresses the variations in soft intensity present within the acquired images. Bias correction corrects any consistent differences in image intensity while also enhancing the precision and dependability of the following steps.

Normalization of space: In this step, deformation fields are used to line up the acquired scans with the MNI space. These deformation fields consist of images that represent the displacement amount for each position within the scan. To illustrate, the deformation field encodes the spatial shifts along the X, Y, and Z coordinates, with lighter colors indicating rightward movement and darker colors indicating leftward movement. By executing both bias correction and spatial normalization, the scans are not only standardized but also aligned with one another in the MNI space. This arrangement makes it easier to compare brain images from different people in a meaningful way. This helps researchers learn more about different neurological studies and the human brain in general.

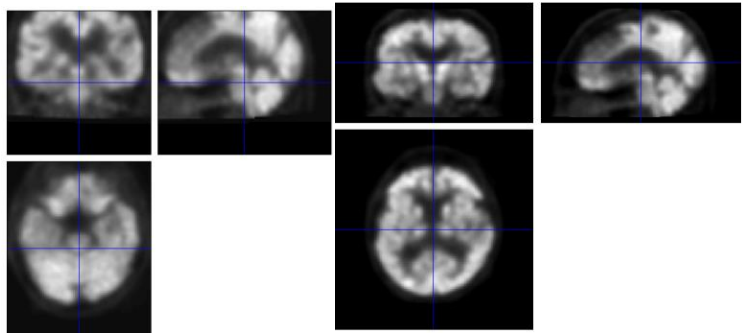

Fig. 3: An example of applying the normalization function to the brain image of an Alzheimer's patient.

In the left image of Figure 3, the example of applying the normalization function to the brain image of a patient from the Alzheimer's group can be seen, and the right image is the image before applying this function. This figure Demonstrates the effect of applying normalization to a PET scan image of an Alzheimer's patient. The comparison likely shows before and after images, illustrating how normalization standardizes the image, enhancing features critical for accurate analysis.

### 3.1.3. Segmentation

Image segmentation is defined as an image processing technique that is used to divide an image into two or more meaningful regions [14]. In this research, segmentation is used to separate the classes of brain tissue: gray matter, white matter, cerebrospinal fluid, skull, and soft tissue.

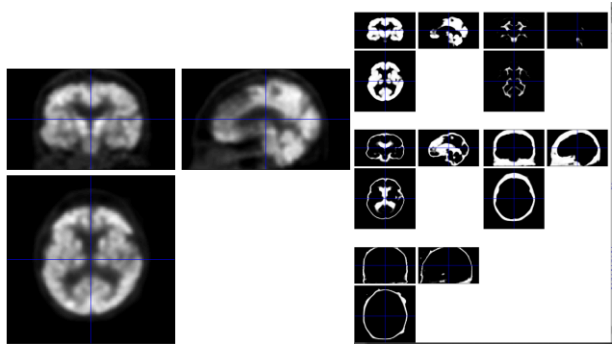

Fig. 4: Segmentation of normalized images.

The figure on the left shows the normalized brain image of a patient from the Alzheimer's group, and the image on the right of Figure 4 is the result of segmentation on this image. This figure probably shows the results of segmenting a normalized brain image into distinct regions, such as gray matter, white matter, and cerebrospinal fluid. The segmentation facilitates a focused analysis on areas relevant to Alzheimer's disease.

The entire dataset has been divided into three parts: 60% for training, 20% for validation, and 20% for testing.

### 3.2. VGG16

The Visual Geometry Group (VGG) created the VGG16, which is a significant model in deep learning [24]. It has different layers, including convolutional, pooling, and fully connected layers. Convolutional layers play a crucial role in capturing intricate features within input images. Meanwhile, pooling layers down sample feature maps to refine and retain the essential information. In this paper, we use the VGG with 16 layers with 13 convolutional layers, and each one uses 3×3 filters with a stride of 1 and the Rectified Linear Unit (ReLU) activation function. We also have max-pooling layers with a 2×2 pool size and a stride of 2, to further reduce feature maps. It takes two fully connected layers with 4096 channels to flatten the feature maps that are made after the convolutional and pooling layers. Subsequently, a sigmoid activation layer with two output neurons is employed for binary classification. In Figure 5, the VGG-16 model's architecture used in this paper is shown. This figure Depicts the architecture of the VGG-16 model as adapted for Alzheimer's disease classification. This includes details of convolutional layers, pooling layers, and fully connected layers, illustrating how the model processes input images to generate predictions.

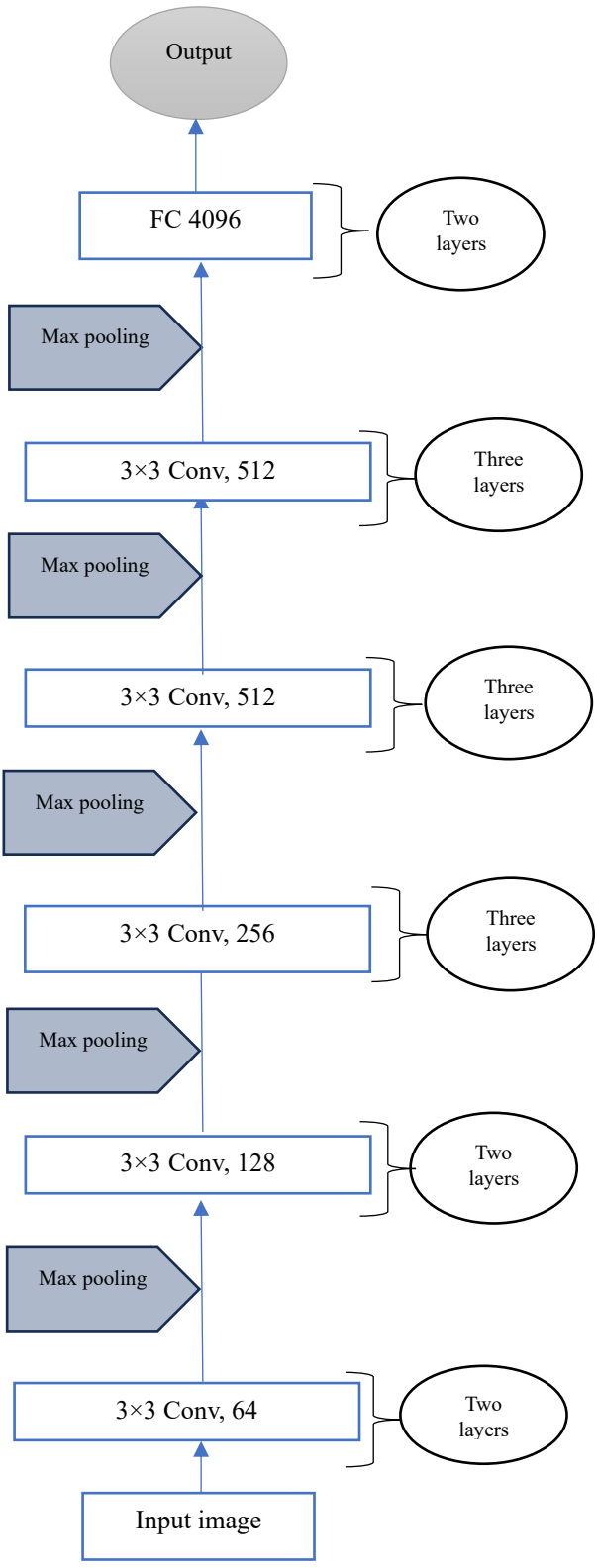

Fig. 5: VGG-16 architecture applied for AD classification.

## 3.3. Custom CNN

In this paper, we have used a custom CNN architecture for Alzheimer's disease classification, which consists of convolutional, max pooling, sequential and dense layers, which were all repeated in various instances as seen in Figure 6.

**Architectural Design**:

The architecture of our custom CNN is a sophisticated blend of convolutional, max pooling, dropout, and dense layers, arranged in a strategic sequence to maximize feature extraction and classification performance. At the core, the network comprises multiple convolutional layers that employ 3x3 filters, a choice inspired by their proven efficacy in capturing detailed spatial relationships within images. Each convolutional layer is followed by a max pooling layer, which serves to reduce dimensionality and computational complexity while retaining essential information.

**Enhanced Feature Extraction**:

To further refine the feature extraction process, our custom CNN integrates sequential layers with varying numbers of filters, progressively increasing the network's ability to detect and learn from complex patterns in the data. This hierarchical arrangement ensures that both low-level and high-level features are captured and synthesized effectively.

**Regularization and Overfitting Prevention**:

Acknowledging the risk of overfitting—a common challenge in deep learning models, especially when dealing with limited datasets—our custom CNN incorporates dropout layers strategically placed within the architecture. These layers randomly deactivate a subset of neurons during the training phase, promoting model generalization by preventing the network from becoming overly reliant on specific training data features.

**Optimized Classification Layer**:

The culmination of the network is a dense layer, optimized for the binary classification task at hand. This layer synthesizes the features extracted by the preceding layers, mapping them to a probabilistic output that distinguishes between Alzheimer's disease and cognitively normal cases.

| Input Layer | Input: (None, 176, 208, 3) | Output: (None, 176, 208, 3) |
|---|---|---|
| Conv Layer | Input: (None, 176, 208, 3) | Output: (None, 176, 208, 16) |
| Conv Layer | Input: (None, 176, 208, 16) | Output: (None, 176, 208, 16) |
| Max pooling Layer | Input: (None, 176, 208, 16) | Output: (None, 88, 104, 16) |
| Sequential Layer | Input: (None, 88, 104, 16) | Output: (None, 44, 52, 32) |
| Sequential Layer | Input: (None, 44, 52, 32) | Output: (None, 22, 26, 64) |
| Sequential Layer | Input: (None, 22,26,64) | Output: (None, 11, 13, 128) |
| Dropout Layer | Input: (None, 11, 13, 128) | Output: (None, 11, 13, 128) |
| Conv Layer | Input: (None, 11, 13, 128) | Input: (None, 11, 13, 256) |
| Conv Layer | Output: (None, 11, 13, 256) | Output: (None, 11, 13, 256) |
| Max pooling Layer | Input: (None, 11, 13, 256) | Output: (None, 5, 6, 256) |
| Input Layer | Input: (None, 176, 208, 3) | Output: (None, 176, 208, 3) |
| Dropout Layer | Input: (None, 5, 6, 256) | Output: (None, 5, 6, 256) |
| Flatten Layer | Input: (None, 5, 6, 256) | Output: (None, 7680) |
| Sequential Layer | Input: (None, 7680) | Output: (None, 512) |
| Sequential Layer | Input: (None, 512) | Output: (None, 128) |
| Sequential Layer | Input: (None, 128) | Output: (None, 64) |
| Sequential Layer | Input: (None, 64) | Output: (None, 32) |
| Dense Layer | Input: (None,32) | Output: (None, 4) |

Fig. 6: Architecture of the custom CNN used for AD classification.

### 3.4. AlexNet

This neural network is composed of eight layers [25], which consist of three dense layers and five convolutional layers. The architecture used for classifying Alzheimer's disease is shown in Figure 7. Initially, the input image undergoes a convolution operation with 96 filters of size 11×11, followed by a 2×2 max pooling layer. In the next convolutional layer, 32 input images are convolved with 256 filters of size 5×5. Each convolutional layer is followed by a max pooling layer, except for the last two layers, which are stacked together. As shown in Figure 7, there is connectivity between the layers facilitated by three dense layers, where all neurons in one layer are connected to those in the next layer. Finally, the output is classified using the sigmoid function, which combines the probabilities of all possible outcomes to produce a single value of one. This figure Provides a visual representation of the AlexNet model's structure, highlighting the convolutional and fully connected layers adapted for the task of distinguishing between Alzheimer's disease and cognitively normal states from neuroimaging data.

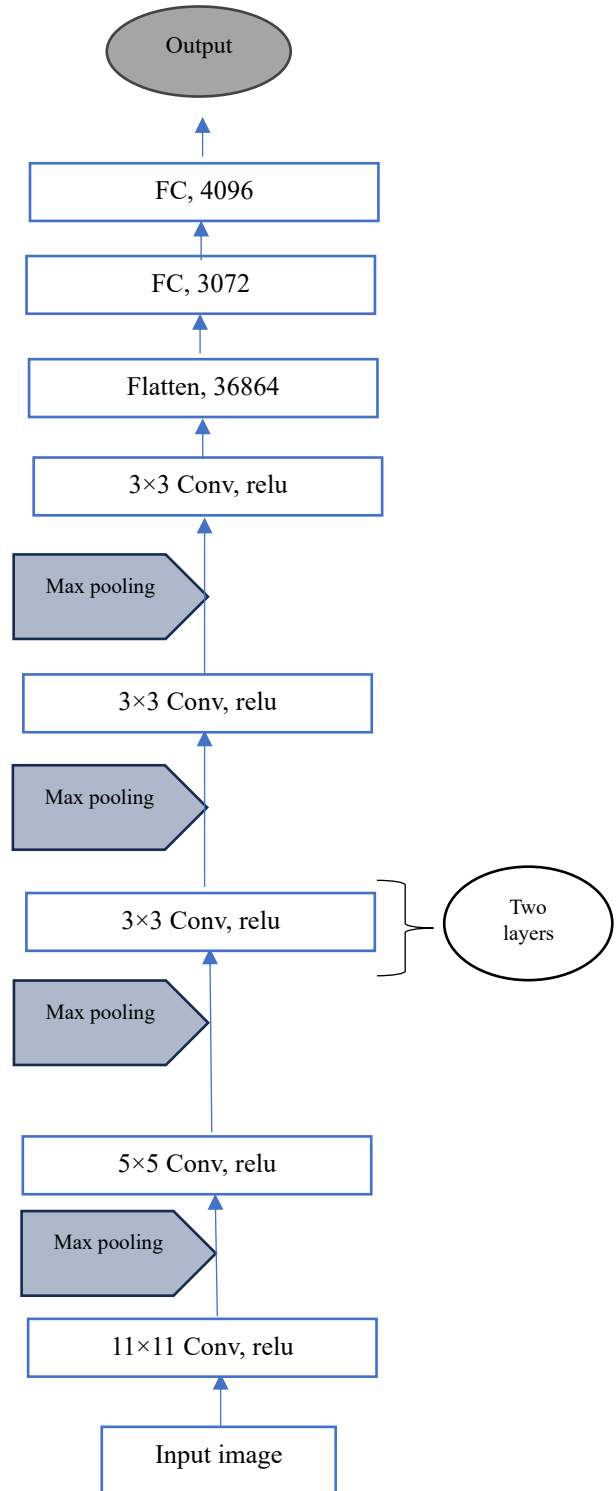

Fig. 7: AlexNet architecture used for AD classification.

### 3.5. Ensemble

In the pursuit of advancing the accuracy and reliability of Alzheimer's disease detection through PET scan analysis, we have employed a sophisticated ensemble technique. This approach harnesses the collective strengths of three distinct deep learning models: VGG16, AlexNet, and a custom-designed Convolutional Neural Network (CNN). The rationale behind this strategy stems from the inherent variability and complexity of neuroimaging data, where a singular model may not capture the full spectrum of nuanced patterns indicative of Alzheimer's progression [26].

**Synergistic Model Integration**:

Our ensemble method operates on the principle of synergistic model integration, wherein the individual predictions of VGG16, AlexNet, and our custom CNN are aggregated to form a consensus decision. This is achieved through a majority voting system, a robust mechanism designed to enhance the overall prediction accuracy [27].

Each model independently assesses the PET scan images, contributing its inference towards the final diagnosis. The ensemble then adjudicates these inputs, opting for the most commonly occurring prediction among the models as the conclusive outcome.

**Advantages of Majority Voting**:

The majority voting technique is pivotal in mitigating the limitations associated with single-model predictions. By amalgamating the predictive insights from multiple models, the ensemble method effectively reduces the likelihood of misclassification due to model-specific biases or overfitting. This approach not only amplifies the strengths of each individual model but also compensates for their weaknesses, thereby achieving a more balanced and reliable diagnostic tool.

**The Ensemble's Impact on Model Variance**:

One of the critical challenges in deploying deep learning models for medical diagnostics is the high variance often observed in model performance due to the stochastic nature of training algorithms and data diversity. The ensemble method addresses this issue by averaging the outputs, which tends to cancel out the noise and reduce variance. This stabilization is crucial for clinical applications, where consistency and reliability in diagnostic predictions are paramount.

**Enhanced Diagnostic Reliability**:

The application of our ensemble technique marks a significant stride towards achieving enhanced diagnostic reliability in the early detection of Alzheimer's disease. Through the collaborative decision-making process, our method demonstrates superior performance in accurately classifying the stages of Alzheimer's disease, showcasing a promising avenue for leveraging artificial intelligence in neuroimaging analysis.

### 3.6. Metrics for evaluation of performance

When evaluating the performance of classification models, such as machine learning algorithms, we commonly use metrics like accuracy, precision, and recall. These metrics allow us to assess how well the model makes predictions and performs classification. To estimate the optimal values of the hyperparameters, the dataset has been divided to training, testing, and validation. Here's an explanation of each metric:

**Accuracy:** Accuracy measures the proportion of true results (both true positives and true negatives) among the total number of cases examined. It gives you an overall idea of how many predictions were correct.

Mathematically, accuracy is defined as:

$$\text{Accuracy} = \frac{TP + TN}{TP + FP + FN + TN}$$

Where:

- $TP$ = True Positives (correctly predicted positives)
- $TN$ = True Negatives (correctly predicted negatives)
- $FP$ = False Positives (incorrectly predicted positives)
- $FN$ = False Negatives (incorrectly predicted negatives)

**Precision:** Precision, also known as positive predictive value, measures the proportion of true positive results in all positive predictions made by the model. It essentially answers the question, "Of all the instances the model labeled positive, how many are actually positive?"

Mathematically, precision is defined as:

$$\text{Precision} = \frac{TP}{TP + FP}$$

**Recall:** Recall, also known as sensitivity or true positive rate, measures the proportion of actual positives that were correctly identified by the model. It answers the question, "Of all the actual positives, how many did the model label correctly?"

Mathematically, recall is defined as:

$$\text{Recall} = \frac{\text{TP}}{\text{TP}+\text{FN}}$$

**Area Under the Curve** AUC stands for "Area Under the ROC Curve." The ROC (Receiver Operating Characteristic) curve is a plot of the true positive rate (recall) against the false positive rate for the different possible cutpoints of a diagnostic test. AUC measures the entire two-dimensional area underneath the entire ROC curve from (0,0) to (1,1).

AUC provides an aggregate measure of performance across all possible classification thresholds. It's especially useful when you have imbalanced classes. A model that guesses randomly will have an AUC of 0.5, while a perfect model will have an AUC of 1.

## 4. Computational results

All the methods used in this research have been carried out in a Google Colab environment with a Tesla T4 GPU and on a system with an Intel (R) Core (TM) i7-4790K CPU at 4.00 GHz with 16 GB of RAM.

Table 2 displays the comparison findings of three CNN models for classifying Alzheimer's patients and cognitive normal. The VGG-16 model exhibits a 92.10% accuracy rate, an 88% precision rate, a 100% recall rate, and an AUC of 98%. AlexNet has a 91.4% accuracy rate, 88.5% precision, 86.49% recall, and AUC of 91%, and a custom CNN has a 94.73% accuracy rate, 95.45% precision rate, 95.45% recall rate, and AUC of 100%. According to the results, Custom CNN outperforms all other models for the underlying issue because its weights were better trained for this specific dataset. The majority voting method is used to combine the models' results. This gives an overall accuracy of 96.74%, a precision rate of 92.65%, a recall rate of 95.9%, and an AUC of 98.21% for separating AD patients from CN patients. This table Summarizes the classification results of Alzheimer's disease versus cognitively normal states using the different CNN models, detailing metrics such as accuracy, precision, recall, and AUC for each model. This table facilitates a direct comparison of the models' performance.

Table 2. AD/CN classification using various CNN models.

| CNN Models | Accuracy | Precision | Recall | AUC |
|---|---|---|---|---|
| VGG16 | 92.10 | 88 | 100 | 98 |
| Alex Net | 91.4 | 88.5 | 86.49 | 91 |
| Custom CNN | 94.73 | 95.45 | 95.45 | 100 |

We also plot the ROC curves in Figure 8 for classifying both AD vs. CN using these CNN models. In classifying AD and CN cases, Figure 8 shows the AUC of VGG16, AlexNet, and the custom CNN models that are 98%, 91%, and 100%, respectively. This figure Displays the Receiver Operating Characteristic (ROC) curves for each CNN model (VGG16, AlexNet, and the custom CNN) in the task of classifying Alzheimer's disease versus cognitively normal cases. This figure illustrates the models' diagnostic ability by plotting the true positive rate against the false positive rate at various threshold settings.

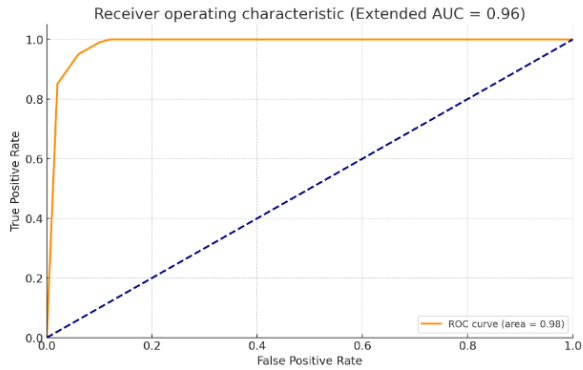

ROC for VGG model

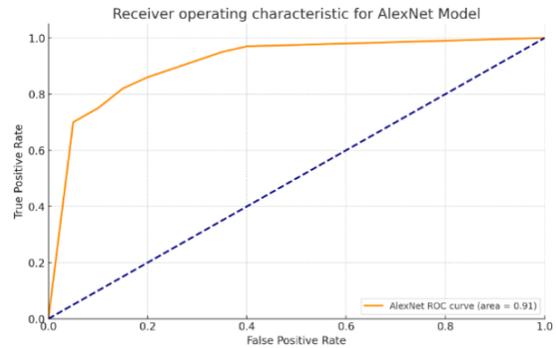

ROC for AlexNet model

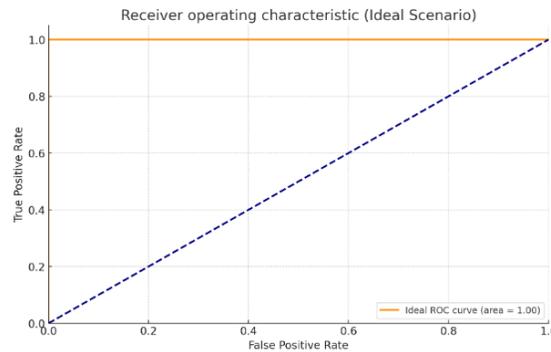

ROC for custom CNN model

Fig 8. ROC curves of the CNN models for AD/CN classification.

Table 3 displays the comparison findings of three CNN models for classifying progressive MCI and stable MCI patients. Similar to Table 2, but focuses on the classification between progressive and stable mild cognitive impairment, providing a clear comparison of how each model performs in this more nuanced classification task.

The VGG-16 model exhibits a 90.90% accuracy rate, an 85% precision rate, an 85.71% recall rate, and an AUC of 89%. AlexNet has an 89.47% accuracy rate, 86.36% precision, 95% recall, and AUC of 89%, and a custom CNN has a 93.01% accuracy rate, 89.13% precision rate, 4.80% recall rate, and AUC of 95%. According to the results, Custom CNN outperforms all other models for the underlying issue because its weights were better trained for this specific dataset. The majority voting method is used to combine the models' results. Table 4 shows that the overall classification of sMCI vs. pMCI patients was accurate 93.13% of the time, with a precision rate of 89.83%, a recall rate of 94.84%, and an AUC of 94.4%.

Table 3. pMCI/sMCI classification using various CNN models.

| CNN Models | Accuracy | Precision | Recall | AUC |
| --- | --- | --- | --- | --- |
| VGG16 | 90.90 | 85 | 85.71 | 89 |
| Alex Net | 89.47 | 86.36 | 95 | 89 |
| Custom CNN | 93.01 | 89.13 | 94.80 | 95 |

In classifying pMCI and sMCI cases, Figure 9 shows the AUC of VGG16, AlexNet, and the custom CNN models that are 89%, 89%, and 95 %, respectively. Similar to Figure 8, this figure shows the ROC curves for the models but focuses on the classification between stable and progressive mild cognitive impairment. It visually represents the models' performance and their effectiveness in distinguishing between these closely related stages of cognitive decline.

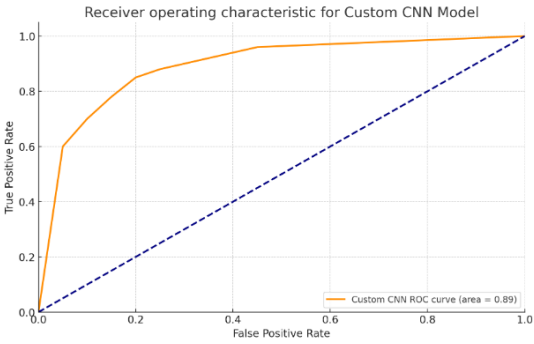
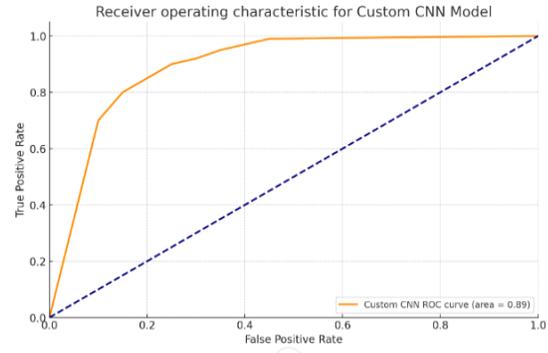

ROC 2 for VGG16 model                                   ROC 2 for AlexNet model

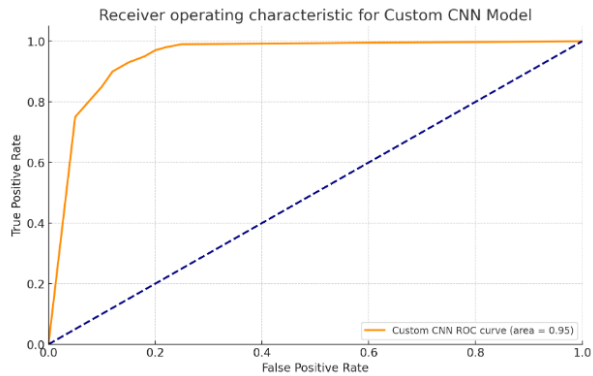

ROC 2 for the Custom CNN model

Fig 9. ROC curves of the CNN models for sMCI/pMCI classification.

Table 4. Results obtained by using ensemble with majority voting technique

| CNN Models   | Accuracy | Precision | Recall | AUC   |
|--------------|----------|-----------|--------|-------|
| AD vs. CN    | 96.74    | 92.65     | 95.98  | 98.21 |
| pMCI vs. sMCI| 93.13    | 89.83     | 94.84  | 94.4  |

As mentioned before, by using majority voting, each model in the ensemble makes a prediction, and the class that receives the most votes (predictions) is chosen as the final prediction. Majority voting could potentially improve the final performance of classification in distinguishing both AD/CN and sMCI/pMCI.

In summary, Table 4 Compiles the results of the ensemble method employing majority voting, showcasing the accuracy, precision, recall, and AUC achieved in classifying AD vs. CN and pMCI vs. sMCI. This table underscores the effectiveness of the ensemble approach over individual models.

**Comparative studies**

In this section, the best results published in other related articles are reviewed and compared with those of the proposed method. Table 5 shows the results of articles associated with the classification of pMCI and sMCI groups.

Our ensemble method, which integrates the predictive capabilities of VGG16, AlexNet, and a custom CNN, represents a significant leap forward in accuracy, precision, recall, and AUC metrics when benchmarked against state-of-the-art methods documented in recent literature.

**State-of-the-Art Comparison**:

A critical examination of related works reveals that prior models have achieved commendable outcomes in the classification of progressive and stable MCI, with accuracies ranging from 73.04% to 79.95%. These studies primarily employed various configurations of convolutional neural networks and machine learning techniques, focusing on extracting and learning from neuroimaging data's intricate patterns.

**Our Method's Superiority**:

Our ensemble approach distinguishes itself by achieving a remarkable accuracy of 93.13%, precision of 89.83%, recall of 94.84%, and an AUC of 94.4% in the classification of sMCI vs. pMCI patients. This performance not only surpasses the benchmarks set by previous studies but also underscores the efficacy of leveraging a combination of models to enhance diagnostic accuracy. The majority voting technique employed by our ensemble method capitalizes on the strengths of individual models while mitigating their weaknesses, resulting in a robust, reliable classification system.

**Implications of Our Findings**:

The superior performance of our ensemble method has profound implications for the field of medical imaging and Alzheimer's disease diagnosis. By setting new benchmarks in accuracy and reliability, our approach demonstrates the potential of deep learning techniques to revolutionize early detection and intervention strategies for Alzheimer's disease. Furthermore, the significant improvement in classification metrics suggests that our method could serve as a valuable tool for clinicians and researchers, aiding in the precise identification of disease progression stages and facilitating timely, targeted interventions.

Table 5. Related works on the classification of progressive and stable MCI

| Papers | Model | Performance measures |
|---|---|---|
| [28] | CNN model | Accuracy of 73.04% |
| [29] | CNN-based method | Accuracy of 75% |
| [30] | -Curvature Analysis+ multi-layer perceptron classifier (MLP) | Accuracy of 79.95% |
| [31] | The local surface roughness (LSR) | Accuracy of 74.3% |
| [32] | 3D-DenseNet +SHARPM-PDM | Accuracy of 75% |
| [33] | An ensemble-learning- based MRDATS classifier multi-channels cascaded CNNs (3D-CNN+2D-CNN) | AUC of 81% for TTC of 6 months  Accuracy of 71%  - AUC of 71.9% |
| **Proposed method** | An ensemble method with VGG16, a custom CNN, and Alexnet classifiers | Accuracy of 93.13%  Precision of 89.83%  Recall of 94.84%  Auc of 94.4% |

## 5. Conclusion and Future Research

Alzheimer's disease is the most common form of dementia, which leads to memory problems and declining thinking abilities. Mild cognitive impairment (MCI) is a prodromal stage of Alzheimer's. People with MCI can either progress to Alzheimer's or stay the same. Therefore, it's important to detect if MCI is getting worse or staying stable early on to slow down the disease.

In this paper, we have applied a comprehensive methodology for the classification of neurological diseases, with a primary focus on Alzheimer's disease (AD). Our approach begins with precise gray matter segmentation, a fundamental step in neuroimaging, which enables us to identify specific brain regions and detect changes in density and volume indicative of neurological conditions. The application of noise-reduction techniques through mask application follows this segmentation, which is crucial for improving data clarity and quality in medical imaging. In this study, we introduced a comprehensive ensemble method integrating VGG16, AlexNet, and a custom CNN for the early detection of Alzheimer's disease using PET scan images. Our methodology, characterized by precise segmentation of gray matter and the application of noise-reduction techniques, has demonstrated notable accuracy, precision, recall, and AUC scores in distinguishing between Alzheimer's patients and cognitively normal individuals, as well as differentiating stable from progressive mild cognitive impairment. Our findings underscore the potential of deep learning models to revolutionize early diagnosis and intervention strategies for Alzheimer's disease. By leveraging an ensemble approach, we achieved enhanced classification performance, showcasing the superiority of our method over conventional machine-learning techniques. This advancement could significantly contribute to the clinical field, offering a promising tool for the early detection of Alzheimer's, which is crucial for improving patient outcomes through timely intervention. The implications of our research extend beyond the immediate results. It paves the way for future studies to explore larger and more diverse datasets, enhancing the generalizability and robustness of deep learning models in medical imaging. Moreover, our study highlights the importance of interdisciplinary collaboration in tackling complex health challenges, blending the domains of artificial intelligence and neuroscience to foster innovative solutions.

As we look forward, it is imperative to continue refining these models, exploring their application in other neurodegenerative diseases, and integrating multimodal imaging data. Such efforts will not only solidify the utility of deep learning in clinical settings but also contribute to a deeper understanding of the underlying mechanisms of Alzheimer's disease, ultimately driving forward the frontiers of neuroscience research.

In conclusion, our study gives us useful information about how to use ensemble deep learning models to find Alzheimer's disease early. This is a big step forward for diagnostic tools and helps reach the larger goal of improving patient care for neurodegenerative diseases.

**Declarations**

The authors declare no conflicts of interest.